\documentclass[twocolumn]{revtex4}
\usepackage{amsmath,amssymb}
\usepackage{graphicx}
\usepackage{dcolumn}
\usepackage[dvips]{epsfig,color}
\usepackage{bm}
\begin{document}
\title{The unquenched quark model\footnote{Talk given at the conference "Dark Matter, Hadron Physics and Fusion Physics", Messina (Italy), September 24-26, 2014}}

\author{E. Santopinto\footnote{email: santopinto@ge.infn.it}} 
\affiliation{Dipartimento di Fisica and INFN, Universit\`a di Genova, via Dodecaneso 33, I-16146 Genova, Italy}
\author{J. Ferretti}
\affiliation{Dipartimento di Fisica and INFN, Universit\`a di Roma "Sapienza", Piazzale A. Moro 5, I-00185 Roma, Italy}

\begin{abstract}
In this contribution, we briefly discuss the results for charmonium and bottomonium spectra with self-energy corrections in the unquenched quark model, due to the coupling to the meson-meson continuum. The UQM formalism can be extended to include also the effects of hybrid mesons, i.e. hybrid loops. Finally, we discuss the results of a calculation of hybrid mesons spectrum in Coulomb Gauge QCD.
\end{abstract}

\maketitle

\section{Introduction}
The quark model can reproduce the behavior of observables such as the spectrum and the magnetic moments, but it neglects pair-creation (or continuum-coupling) effects.
Above threshold, this coupling leads to strong decays; below threshold, it leads to virtual $q \bar q - q \bar q$ components in the meson wave function and shifts of the physical mass with respect to the bare mass.  
The unquenching of the quark model for mesons is a way to take these components into account.

Pioneering work on the unquenching of meson quark models was done by Van Beveren, Dullemond and Rupp using a t-matrix approach \cite{vanBeveren:1979bd,vanBeveren:1986ea}, while T\"ornqvist and collaborators \cite{Ono:1983rd,Tornqvist} used their unitarized QM.
These methods were used (with a few variations) by several authors to study the influence of the meson-meson continuum on meson observables. 
We mention, as an example, the study of the scalar meson nonet ($a_0$, $f_0$, etc.) of Ref. \cite{vanBeveren:1986ea,Tornqvist:1995kr} in which the loop contributions are given by the hadronic intermediate states that each meson can access. It is via these hadronic loops that the bare states become "dressed" and  the hadronic loop contributions totally dominante the dynamics of the process. A very similar approach was developed by Pennington in Ref. \cite{Pennington:2002}, in which the dynamical generation of the scalar mesons by initially inserting only one "bare seed", was investigated. The study of Ref. \cite{Geiger:1989yc} demonstrates that the effects of the $q \bar q$ sea pairs in meson spectroscopy is simply a renormalization of the meson string tension. 
Eichten {\it et al.} explored the influence of the open-charm channels on the charmonium properties, using the Cornell coupled-channel model \cite{Eichten:1974af} to assess departures from the single-channel  potential-model expectations.
  
In this contribution, we discuss some of the latest applications of the UQM to the study of meson observables. Finally, we discuss the spectroscopy of hybrid mesons in Coulomb Gauge QCD.	

\section{UQM }
\subsection{Formalism} 
\label{Sec:formalism}
In the unquenched quark model for mesons \cite{bottomonium,charmonium,Ferretti:2013vua,Ferretti:2014xqa}, the meson wave function is made up the valence $q \bar q$ configuration plus a sum over the continuum components as
\begin{eqnarray} 
	\label{eqn:Psi-A}
	\mid \psi_A \rangle &=& {\cal N} \left[ \mid A \rangle 
	+ \sum_{BC \ell J} \int d \vec{K} \, k^2 dk \, \mid BC \ell J;\vec{K} k \rangle \right.
	\nonumber\\
	&& \hspace{2cm} \left.  \frac{ \langle BC \ell J;\vec{K} k \mid T^{\dagger} \mid A \rangle } 
	{E_a - E_b - E_c} \right] ~, 
\end{eqnarray}
where $T^{\dagger}$ stands for the $^{3}P_0$ quark-antiquark pair-creation operator \cite{bottomonium,charmonium,Ferretti:2013vua,Ferretti:2014xqa}, $A$ is the meson, $B$ and $C$ represent the intermediate virtual mesons, $E_a$, $E_b$ and $E_c$ are the corresponding energies, $k$ and $\ell$ the relative radial momentum and orbital angular momentum between $B$ and $C$ and $\vec{J} = \vec{J}_b + \vec{J}_c + \vec{\ell}$ is the total angular momentum. 
It is worthwhile noting that in Refs. \cite{bottomonium,charmonium,Ferretti:2013vua,Ferretti:2014xqa,Kalashnikova:2005ui}, the constant pair-creation strength in the operator (\ref{eqn:Psi-A}) was substituted with an effective one, to suppress unphysical heavy quark pair-creation. 

In the UQM \cite{bottomonium,charmonium,Ferretti:2013vua,Ferretti:2014xqa}, the matrix elements of an observable $\hat O$ can be calculated as
\begin{equation}
	O = \left\langle \psi_A \right| \hat O \left| \psi_A \right\rangle \mbox{ }, 
\end{equation}
where $\left| \psi_A \right\rangle$ is the state of Eq. (\ref{eqn:Psi-A}). 
The result will receive a contribution from the valence part and one from the continuum component, which is absent in naive QM calculations.  

\subsection{$c \bar c$ and $b \bar b$ spectra with self-energy corrections in the UQM}
In Refs. \cite{bottomonium,charmonium,Ferretti:2013vua,Ferretti:2014xqa}, the method was used to calculate the $c \bar c$ and $b \bar b$ spectra with self-energy corrections, due to continuum coupling effects. In the UQM, the physical mass of a meson, 
\begin{equation}
	\label{eqn:self-trascendental}
	M_a = E_a + \Sigma(E_a)  \mbox{ },
\end{equation}
is given by the sum of two terms: a bare energy, $E_a$, calculated within a potential model 
\cite{Godfrey:1985xj}, and a self energy correction, 
\begin{equation}
	\label{eqn:self-a}
	\Sigma(E_a) = \sum_{BC\ell J} \int_0^{\infty} k^2 dk \mbox{ } \frac{\left| M_{A \rightarrow BC}(k) \right|^2}{E_a - E_b - E_c}  \mbox{ },
\end{equation}
computed within the UQM formalism. 

Our results for the self energies of charmonia \cite{charmonium,Ferretti:2014xqa} and bottomonia \cite{bottomonium,Ferretti:2013vua,Ferretti:2014xqa} show that the pair-creation effects on the spectrum of heavy mesons are quite small. Specifically for charmonium and bottomonium states, they are of the order of $2 - 6\%$ and $1 \%$, respectively. 
The relative mass shifts, i.e. the difference between the self energies of two meson states, are in the order of a few tens of MeV. 
However, as QM's can predict the meson masses with relatively high precision in the heavy quark sector, even these corrections can become significant.
These results are particularly interesting in the case of states close to an open-flavor decay threshold, like the $X(3872)$ and $\chi_b(3P)$ mesons.
For example, in our picture the $X(3872)$ can be interpreted as a $c \bar c$ core, working as a "seed" [the $\chi_{c1}(2^3P_1)$], plus the virtual meson-meson continuum. In Ref. \cite{Ferretti:2014xqa}, we showed that the probability to find the $X(3872)$ in its core or continuum components is approximately $45\%$ and $55\%$, respectively.

\subsection{Loops of hybrid mesons}
The wave function of a meson can be written as
\begin{equation}
	\label{eqn:higher-Fock}
	\left| \Psi \right\rangle = \left| q \bar q \right\rangle + \left| q \bar q q \bar q \right\rangle
	+ \left| q \bar q g \right\rangle + ... \mbox{ },
\end{equation}
where $\left| q \bar q \right\rangle$ is the quark-antiquark component, $\left| q \bar q q \bar q \right\rangle$ the tetraquark or molecular component and $\left| q \bar q g \right\rangle$ the quark-antiquark-gluon (hybrid) component.
In the QM, conventional mesons are in general described by the $\left| q \bar q \right\rangle$ valence component. Nevertheless, there are also attempts to accommodate exotic states as $\left| q \bar q q \bar q \right\rangle$ or $\left| q \bar q g \right\rangle$ states.

In the QM formalism, $\left| q \bar q \right\rangle$ and $\left| q \bar q g \right\rangle$ mesons are described by the non relativistic Hamiltonian \cite{Isgur:1984bm}
\begin{equation} 
	\label{eqn:Hstring}
	H^\nu = - \frac{1}{2 \mu} \frac{\partial^2}{\partial r^2} + \frac{\ell (\ell+1)}{2 \mu r^2} + E^\nu(r)  \mbox{ }, 
\end{equation}
where 
\begin{equation}
  \label{eqn:Enu}
	E^\nu(r) = - \frac{\tau}{r} + \beta r + \frac{\nu \pi}{r} + C  \mbox{ },  
\end{equation}
is the potential describing $q \bar q$ mesons ($\nu = 0$), the first hybrid surface ($\nu = 1$), the second hybrid surface ($\nu = 2$), and so on.
The symbols $\ell$ and $\mu$ in Eq. (\ref{eqn:Enu}) stand for the orbital angular momentum of the state and the reduced mass of the $q \bar q$ system, respectively.

In the UQM formalism, the wave function of a meson can be written as the superposition of the components of Eq. (\ref{eqn:higher-Fock}). 
Up to now, we have considered in our calculations \cite{bottomonium,charmonium,Ferretti:2013vua,Ferretti:2014xqa} only the first two terms, $\left| q \bar q \right\rangle$ and $\left| q \bar q q \bar q \right\rangle$.
They are the most important for ground state and lower-lying mesons, close to the first open-flavor decay thresholds. 
Nevertheless, at higher energies, the effects of ground state open-flavor meson loops, like $D \bar D$ or $D \bar D^*$ in the $c \bar c$ sector (see Ref. \cite{charmonium}), become less important.
In the $c \bar c$ sector, the first hybrid $c \bar c g$ mesons lie at energies of 4.2 GeV, approximately.
Thus, above 4 GeV, the introduction of hybrid loops could be crucial to understanding the higher lying mesons' structure and spectrum. 

In the QM formalism, the coupling between a hybrid meson $\mathcal H$ and a quarkonium state $Q$ is given by \cite{LeYaouanc:1984gh}
\begin{equation}
	\label{eqn:Hibrid-coupling}
	\left\langle \mathcal H \right| V \left| Q \right\rangle = \left\langle q \bar q g \right| V \left| q \bar q \right\rangle \mbox{ },
\end{equation}
where $V$ is an interaction that annihilates the constituent gluon $g$.
The coupling of Eq. (\ref{eqn:Hibrid-coupling}) can be used to calculate the contribution of hybrid loops to the self-energy of higher-lying mesons \cite{FS01}.

\begin{figure}[htbp]
\centerline{%
\includegraphics[width=7cm,angle=270]{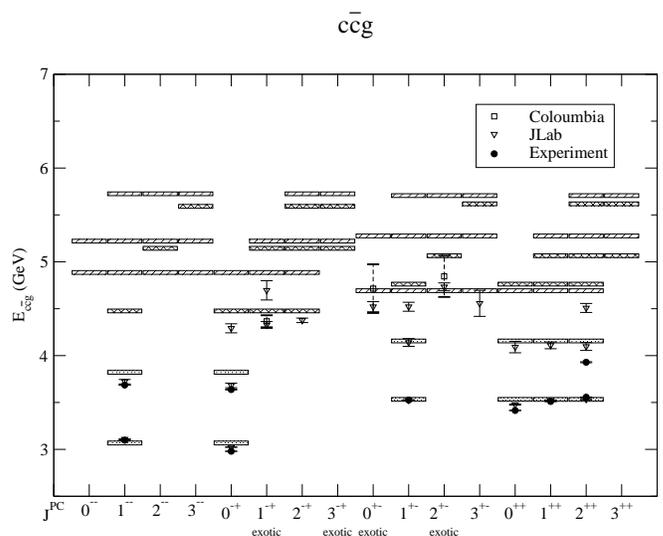}}
\caption{Charmonium (solid boxes) and charmonium hybrid spectrum compared with data (where available) or lattice computations. Single dashed boxes are the $c{\bar c}g$ hybrids dominated by the $P$-wave quarks, all other have the $Q{\bar Q}$ pair in the relative $S$-wave orbital. Picture from Ref. \cite{Guo:2008yz}; APS copyright.}
\label{Fig:F1}
\end{figure}

\begin{figure}[htbp]
\centerline{%
\includegraphics[width=7cm,angle=270]{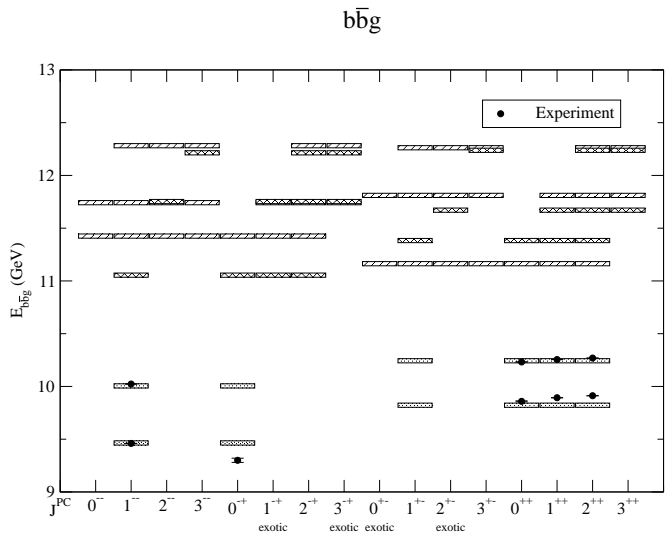}}
\caption{Same as Fig. \ref{Fig:F1} for bottomonium. Picture from Ref. \cite{Guo:2008yz}; APS copyright.}
\label{Fig:F2}
\end{figure}

\section{Hybrid mesons in Coulomb Gauge QCD}
Hybrid states have been studied in several models \cite{Isgur:1984bm,Horn:1977rq,Szczepaniak:1995cw,Simonov:2000ky,Wang:2006ida,Buisseret:2006wc}, including lattice QCD calculations (for example, see Ref. \cite{Morningstar:1999rf}). 
The most commonly studied hybrids are composed of a quark, an antiquark and a gluon.
Particular interest has grown because of the recent discovery of several new states, in particular in the $c \bar c$ sector, probably including a hybrid resonance, the $Y(4260)$, with $1^{--}$ quantum numbers. This meson was discovered by BaBar in ISR production $J/\Psi \pi^+ \pi^-$ \cite{Aubert:2005rm} and then confirmed by CLEO \cite{CLEO:2006aa0} and BELLE \cite{Pakhlova:2010ek}.
The observation of $Y (4260)$ by ISR strongly suggests that it is a vector meson. In direct production, CLEO also observed the $Y (4260)$ decaying into $J/\Psi \pi^0\pi^0$ and $J/\Psi K^+ K^-$ \cite{CLEO:2006aa}. Although there may be not a large amount of data to allow an interpretation of the $Y (4260)$ based on its decay properties, the present state of hadronic theory is not up to the task.
Since no conventional charmonium vectors are expected at this mass, the $Y (4260)$ may be dominantly a hybrid meson.

Conventional heavy quarkonia are well described by non-relativistic QCD, thus one can expect that hybrids containing heavy quarks can be treated similarly, considering gluonic excitations in presence of slow quarks. Moreover, in physical gauge, the dynamical gluons can be separated from the instantaneous Coulomb-type forces that act between color charges, thus while the non-abelian Coulomb potential is expected to be responsible for binding and confinement, the remaining, transverse gluon excitations could bring contribution to the spectrum.

In Refs. \cite{Guo:2007sm,Guo:2008yz}, hybrid mesons were studied in a variational approach to QCD in the Coulomb gauge. In this approach, a confining linear potential emerged from the Dyson-Schwinger equations, at least at the hadronic scale. In a first stage, this potential was used to calculate the spectrum of the gluelump, which is an idealized system defined as gluonic excitations bounded to a static, localized color octet source, such as, for example, a very heavy quark and antiquark \cite{Guo:2007sm}. The next step was to introduce the quark-antiquark dynamics to calculate the spectrum of heavy hybrid mesons \cite{Guo:2008yz}. See Figs. \ref{Fig:F1} and \ref{Fig:F2}, where the results are compared with data (where available) or lattice calculations.

It is worthwhile noting that the lowest mass charmonium hybrid multiplet was predicted to lie at an energy of 4.476 GeV and to be composed by states with $J^{PC} = 1^{--}$; $(0; 1; 2)^{-+}$ \cite{Guo:2008yz}; the multiplet includes also an exotic state, with $J^{PC} = 1^{-+}$. This four state hybrid multiplet was also recently identified in lattice simulations \cite{Dudek:2011tt}. 
In the bottomonium case, the lowest mass hybrid multiplet lies at an energy of 11.055 GeV and is composed, as in the case of $c \bar c g$, by states with $J^{PC} = 1^{--}$, $0^{-+}$, $1^{-+}$ and $2^{-+}$. The $1^{--}$ state could be identified with the resonance $Y_b(10888)$ observed by Belle in $e^+e^-$ annihilation \cite{Chen:2010}.


\begin{thebibliography}{}
	
\bibitem{vanBeveren:1979bd} 
  E.~van Beveren, C.~Dullemond and G.~Rupp,
  Phys.\ Rev.\ D {\bf 21}, 772 (1980)
  [Erratum-ibid.\ D {\bf 22}, 787 (1980)].

\bibitem{vanBeveren:1986ea} 
  E.~van Beveren, T.~A.~Rijken, K.~Metzger, C.~Dullemond, G.~Rupp and J.~E.~Ribeiro,
  Z.\ Phys.\ C {\bf 30}, 615 (1986).
	
  \bibitem{Ono:1983rd}
  S.~Ono and N.~A.~T\"ornqvist,
  Z.\ Phys.\ C {\bf 23}, 59 (1984);
  K.~Heikkila, S.~Ono and N.~A.~T\"ornqvist,
  Phys.\ Rev.\ D {\bf 29}, 110 (1984)
  [Erratum-ibid.\ {\bf 29}, 2136 (1984)];
  S.~Ono, A.~I.~Sanda and N.~A.~T\"ornqvist,
  Phys.\ Rev.\ D {\bf 34}, 186 (1986).  
		
\bibitem{Tornqvist}
  N.~A.~T\"ornqvist and P.~Zenczykowski,
  Phys.\ Rev.\  D {\bf 29}, 2139 (1984);
  Z.\ Phys.\  C {\bf 30}, 83 (1986);
  P.~Zenczykowski,
  Annals Phys.\  {\bf 169}, 453 (1986).	
	
\bibitem{Tornqvist:1995kr} 
  N.~A.~Tornqvist,
  Z.\ Phys.\ C {\bf 68}, 647 (1995).
	
\bibitem{Pennington:2002} 
  M.~Boglione and M. R. ~Pennington,
  Phys.\ Rev.\ D {\bf 65}, 114010 (2002).	

\bibitem{Geiger:1989yc} 
  P.~Geiger and N.~Isgur,
  Phys.\ Rev.\ D {\bf 41}, 1595 (1990).	   
  
\bibitem{Eichten:1974af}  
  E.~Eichten, K.~Gottfried, T.~Kinoshita, J.~B.~Kogut, K.~D.~Lane and T.~-M.~Yan,
  Phys.\ Rev.\ Lett.\  {\bf 34}, 369 (1975);
  E.~Eichten, K.~Gottfried, T.~Kinoshita, K.~D.~Lane and T.~-M.~Yan,
  Phys.\ Rev.\ D {\bf 17}, 3090 (1978);
  {\bf 21}, 203 (1980).	
		
\bibitem{bottomonium}
  J.~Ferretti, G.~Galat\'a, E.~Santopinto and A.~Vassallo,
  Phys.\ Rev.\ C {\bf 86}, 015204 (2012).  
	
\bibitem{charmonium}  
   J.~Ferretti, G.~Galat\'a and E.~Santopinto,
  Phys.\ Rev.\ C {\bf 88}, 015207 (2013).	
	
\bibitem{Ferretti:2013vua}   
  J.~Ferretti and E.~Santopinto,
  Phys.\ Rev.\  D {\bf 90}, 094022 (2014).	
  
\bibitem{Ferretti:2014xqa} 
  J.~Ferretti, G.~Galat\'a and E.~Santopinto,
  Phys.\ Rev.\  D {\bf 90}, 054010 (2014).	
	
\bibitem{Godfrey:1985xj}
  S.~Godfrey and N.~Isgur,
  Phys.\ Rev.\ D \textbf{ 32}, 189 (1985).
	
\bibitem{Kalashnikova:2005ui}
  Y.~S.~Kalashnikova,
  Phys.\ Rev.\ D {\bf 72}, 034010 (2005).						
					
\bibitem{Isgur:1984bm} 
  N.~Isgur and J.~E.~Paton,
  Phys.\ Rev.\ D {\bf 31}, 2910 (1985).	
	
\bibitem{LeYaouanc:1984gh} 
  A.~Le Yaouanc, L.~Oliver, O.~Pene, J.~C.~Raynal and S.~Ono,
  Z.\ Phys.\ C {\bf 28}, 309 (1985).	
						
\bibitem{FS01} 
  J.~Ferretti and E.~Santopinto,
  unpublished.						
									
\bibitem{Horn:1977rq} 
  D.~Horn and J.~Mandula,
  Phys.\ Rev.\ D {\bf 17}, 898 (1978).
	
\bibitem{Szczepaniak:1995cw} 
  A.~Szczepaniak, E.~S.~Swanson, C.~R.~Ji and S.~R.~Cotanch,
  Phys.\ Rev.\ Lett.\  {\bf 76}, 2011 (1996).	
										
\bibitem{Simonov:2000ky} 
  Y.~A.~Simonov,
  Nucl.\ Phys.\ B {\bf 592}, 350 (2001).
	
\bibitem{Wang:2006ida} 
  Z.~G.~Wang and S.~L.~Wan,
  Phys.\ Rev.\ D {\bf 74}, 014017 (2006).	
											
\bibitem{Buisseret:2006wc} 
  F.~Buisseret and C.~Semay,
  Phys.\ Rev.\ D {\bf 74}, 114018 (2006).
	
\bibitem{Morningstar:1999rf} 
  C.~J.~Morningstar and M.~J.~Peardon,
  Phys.\ Rev.\ D {\bf 60}, 034509 (1999).	
	
\bibitem{Aubert:2005rm} 
  B.~Aubert {\it et al.}  [BaBar Collaboration],
  Phys.\ Rev.\ Lett.\  {\bf 95}, 142001 (2005).
	
\bibitem{CLEO:2006aa0}
 Q.~He {\it et al.}  [CLEO Collaboration],
  Phys.\ Rev.\ D {\bf 74}, 091104 (2006)
	
		
\bibitem{Pakhlova:2010ek} 
  G.~Pakhlova {\it et al.}  [Belle Collaboration],
  Phys.\ Rev.\ D {\bf 83}, 011101 (2011).		
		
\bibitem{CLEO:2006aa} 
  T. E.~Coan  {\it et al.}  [CLEO Collaboration],
  Phys.\ Rev.\ Lett.\  {\bf 96}, 162003 (2006).
		
\bibitem{Guo:2007sm} 
  P.~Guo, A.~P.~Szczepaniak, G.~Galata, A.~Vassallo and E.~Santopinto,
  Phys.\ Rev.\ D {\bf 77}, 056005 (2008).
			
\bibitem{Guo:2008yz} 
  P.~Guo, A.~P.~Szczepaniak, G.~Galata, A.~Vassallo and E.~Santopinto,
  Phys.\ Rev.\ D {\bf 78}, 056003 (2008).			
		
\bibitem{Dudek:2011tt} 
  J.~J.~Dudek, R.~G.~Edwards, B.~Joo, M.~J.~Peardon, D.~G.~Richards and C.~E.~Thomas,
  Phys.\ Rev.\ D {\bf 83}, 111502 (2011).		
	
\bibitem{Chen:2010} 
  K.-F.~Chen {\it et al.}  [Belle Collaboration],
  Phys.\ Rev.\ D {\bf 82}, 091106 (2010).
	
\end{thebibliography}
\end{document}